\documentclass{svproc}
\usepackage{graphicx}%
\usepackage{multirow}%
\usepackage{amsmath,amssymb,amsfonts}%
\usepackage{mathrsfs}%
\usepackage[title]{appendix}%
\usepackage[table]{xcolor}%
\usepackage{textcomp}%
\usepackage{manyfoot}%
\usepackage{booktabs}%
\usepackage{algorithm}%
\usepackage{algorithmicx}%
\usepackage{algpseudocode}%
\usepackage{listings}%
\usepackage{subcaption}%
\usepackage{url}

\def\orcidID#1{\unskip$^{[#1]}$} 

\usepackage{hyperref}

\usepackage{tikz}
\newcommand{\tikzcircle}[3][0.95ex]{%
    \definecolor{fillcolor}{HTML}{#2}%
    \definecolor{bordercolor}{HTML}{#3}%
    \tikz[baseline=-0.75ex] \draw[fill=fillcolor, draw=bordercolor, radius=#1] (0,0) circle;%
}
\definecolor{tablecol1}{HTML}{4c78a8}
\begin{document}
\mainmatter              
\title{Inside Alameda Research: A Multi-Token Network Analysis}
%
\titlerunning{Inside Alameda Research: A Multi-Token Network Analysis}  
%
\author{Célestin Coquidé\inst{1}\orcidID{0000-0001-8546-6587} \and
Rémy Cazabet\inst{1}\orcidID{0000-0002-9429-3865} \and Natkamon Tovanich\inst{2}\orcidID{0000-0001-9680-9282}}
\authorrunning{Coquidé, Cazabet, and Tovanich} 

\institute{Univ Lyon, UCBL, CNRS, LIRIS, UMR 5205, 69622 Villeurbanne, France
\email{celestin.coquide@liris.cnrs.fr}, \email{remy.cazabet@gmail.com},\\
\and
CREST, CNRS, École Polytechnique, Institute Polytechnique de Paris,\\
91120 Palaiseau, France
\email{natkamon.tovanich@polytechnique.edu}}

\maketitle              

\begin{abstract}
We analyze the token transfer network on Ethereum, focusing on accounts associated with Alameda Research, a cryptocurrency trading firm implicated in the misuse of FTX customer funds. Using a multi-token network representation, we examine node centralities and the network backbone to identify critical accounts, tokens, and activity groups. The temporal evolution of Alameda accounts reveals shifts in token accumulation and distribution patterns leading up to its bankruptcy in November 2022. Through network analysis, our work offers insights into the activities and dynamics that shape the DeFi ecosystem.
\end{abstract}
%



\setcounter{footnote}{0}

\section{Introduction}
\label{sec:introduction}

Ethereum, introduced in 2015, is the leading blockchain platform for smart contracts---self-executing computer programs operate on a decentralized network~\cite{buterin2014next}. Unlike Bitcoin, which primarily serves as a digital currency, smart contracts enable various applications, including tokenization, where digital tokens can be created, destroyed, and transferred within the blockchain \cite{wang2021sok}.

This capability has given rise to Decentralized Finance (DeFi), a rapidly growing sector that replicates and extends traditional financial services in a decentralized environment \cite{schar2021decentralized}. DeFi consists of protocols allowing users to perform operations like exchanging, lending, borrowing, and arbitrage without centralized intermediaries. It revolutionizes the financial landscape by providing users with autonomy and flexibility.

Although the DeFi ecosystem has been successful and grown exponentially, this rapid expansion has also exposed vulnerabilities within the system, as exemplified by the collapse of major entities like FTX, a prominent cryptocurrency exchange company \cite{conlon2024contagion,fu2023ftx}. The fall of FTX sent shockwaves through the crypto world, leading to widespread scrutiny of its practices. While extensive analyses have been conducted on FTX itself, less attention has been paid to its sister company, Alameda Research, which is responsible for managing significant portions of customer funds.

Alameda Research is a quantitative cryptocurrency trading company co-founded by the same individual as FTX. Due to the deep interconnection between the two entities, FTX reportedly lent out a substantial portion of its customers' assets, raising significant questions about both companies' internal operations and the movement of tokens within their networks.
Understanding the dynamics of these token transfers is crucial, as it sheds light on the underlying mechanisms that may have contributed to the downfall. However, analyzing DeFi entities like Alameda is inherently challenging due to the complexity of interacting tokens, diverse protocols, and the opaque nature of transactions on the blockchain.



This paper aims to bridge the gap in our understanding by conducting an analysis of the token transfer networks associated with Alameda Research.
We seek to uncover patterns and anomalies within these transfers and analyze the activities that occurred before and during the collapse.
Our work not only highlights the specific case of Alameda but also underscores the broader challenges of analyzing DeFi ecosystems, where the interplay of various tokens and protocols can obscure critical information.

\vspace{-2mm}
\section{Related Work}
\label{sec:related_work}

Various networks can be derived from Ethereum blockchain data, such as transaction networks \cite{bai2021evolution,chen2020understanding,kondor2021rich,lee2020measurements,zhao2021temporal}, ERC-20 token networks \cite{chen2020traveling,loporchio2023analysis,morales2023patterns,victor2019measuring,zhu2024data}, and non-fungible token (NFT) networks \cite{alizadeh2023network,casale2021networks,la2023game,nadini2021mapping,von2022nft}. Early studies investigated the structural properties of Ethereum's transaction networks \cite{chen2020understanding,lee2020measurements}, highlighting the importance of graph-based methods for understanding entity interactions. Temporal studies revealed network dynamics and the ``rich-get-richer'' effect \cite{bai2021evolution,kondor2021rich,zhao2021temporal}.

Much research on Ethereum token networks examined the topology and interactions of ERC-20 tokens \cite{chen2020traveling,loporchio2023analysis,victor2019measuring}. Morales et al. studied heterogeneity in user roles
and behaviors regarding transaction diversity and activity. \cite{morales2023patterns}. Zhu et al. identified anomaly patterns in the USDC token network during key events like the LunaTerra collapse using k-core decomposition and graph motifs \cite{zhu2024data}.

The unique features of the NFT trading network have also drawn attention. Alizadeh et al. and Casale-Brunet et al. conducted a graph analysis of the NFT network \cite{alizadeh2023network,casale2021networks}. Nadini et al. further mapped the NFT revolution by analyzing market trends, trade networks, and visual features \cite{nadini2021mapping}. Wash trading detection using cycles in token transfer networks has also been explored \cite{la2023game,von2022nft}.

While mainstream media and a few papers, such as \cite{conlon2024contagion,fu2023ftx}, have investigated the mechanisms behind the FTX collapse, there has been limited analysis of the activities within Alameda's network. Our work addresses this gap by characterizing key entities and analyzing interactions between user accounts, contracts, and tokens. This approach provides a clearer understanding of participant behaviors and the flow of value within the DeFi ecosystem.

\vspace{-2mm}
\section{Dataset}
\label{sec:dataset}

In the blockchain, tokens can be freely transferred between accounts and protocols, forming a massive and complex network of transactions.
Digital tokens represent a unit of value or asset that can be traded or transferred on a blockchain.
Ethereum accounts are divided into two types: \emph{externally owned accounts (EOAs)} and \emph{contract accounts (CAs)}. EOAs are controlled by private keys and typically belong to users, while CAs are associated with smart contracts, which represent protocols or execute automated functions on the blockchain.

\paragraph{\textbf{List of fund accounts:}}
We obtained a list of labeled accounts from Etherscan, focusing on those tagged as ``fund'' accounts \footnote{\url{https://etherscan.io/accounts/label/fund}}. These fund accounts represent entities on the blockchain that engage in cryptocurrency exchanges. Our dataset includes 65 such accounts, with 27 explicitly linked to Alameda Research, labeled as Alameda Research 1, 2, ..., 27, with the exceptions of accounts 18 and 21, which are labeled as \emph{Binance Deposit} and \emph{WBTC Merchant Deposit} accounts. By collecting token transfers related to these ``fund'' accounts, we aimed to analyze the broader network of financial activities within the blockchain ecosystem, particularly those associated with significant entities like Alameda Research. Including other ``fund'' accounts allows us to contextualize Alameda Research's actions in the broader landscape of cryptocurrency exchanges.


\paragraph{\textbf{Token transfer data:}}
To obtain the token transfer data, we used Alchemy's Transfer API\footnote{\url{https://docs.alchemy.com/reference/transfers-api-quickstart}}, which returns a complete list of token transfer history for any transaction, address, or block. We collect token transfers from and to each fund account, updated until the end of March 2023.
Each transfer records \emph{block number, timestamp, transaction hash, from address, to address, token, and value}.


Specifically, the ``from'' and ``to'' addresses are represented as the last 20 bytes of the Keccak-256 hash of an ECDSA public key. These addresses are \emph{pseudonymous}, meaning the identity of an owner is not directly tied to the address unless disclosed by the owner has publicly disclosed their address or it has been linked to their identity through other means.

\paragraph{\textbf{Labeled addresses}}
To enhance our analysis, we scraped address webpages from Etherscan to collect available public name tags that indicate the identity or role of each address\footnote{ \url{https://info.etherscan.com/public-name-tags-labels/}}. We also gathered information on whether each address is a user account (EOA) or a contract account (CA). Out of the 68,907 distinct addresses in our token transfer dataset, we identified 3,627 name tags (5.26\%).

In the blockchain, a single entity can control one or multiple addresses. For instance, \emph{Alameda Research}, the focus of our study, controls 27 different accounts (as known and tagged by Etherscan), each used for different activities. Similarly, DeFi protocols like \emph{Compound} or \emph{Uniswap} are operated through multiple contracts, each representing a different liquidity pool. We noticed that the name tag from Etherscan uses a ``:'' delimiter to separate the entity and its contract name. For instance, ``Compound: cETH Token'' and ``Compound: cDAI Token'' are liquidity pools within the \emph{Compound} protocol, while ``Uniswap V3: YFI'' and ``Uniswap V3: 1INCH-USDC'' belong to the \emph{Uniswap V3} protocol. Using this pattern, we can group addresses that belong to the same entity, resulting in a total of 1,082 entities. Untagged addresses were treated as separate entities.


\paragraph{\textbf{Filtering out spam tokens}}
Another issue we encountered in handling this dataset is the proliferation of spam tokens designed to lure traders into purchasing them. These tokens often target well-known addresses by distributing airdrops to boost their perceived legitimacy. Despite having no actual value, these tokens can appear similar to legitimate coins or even replicate well-known brand names. These tokens potentially contaminate our data and lead to inaccurate network statistics.
To ensure the accuracy and reliability of our dataset, we manually verified the top 500 tokens based on the number of transactions in our token transfer dataset. Additionally, we collected a list of verified tokens from Etherscan\footnote{\url{https://etherscan.io/tokens}} and used it to filter out transactions involving unverified tokens.

After data preprocessing steps, the resulting token transfer dataset consists of 1,473,736 transfers (92\% of the original token transfer dataset) involving 65,403 addresses (95\%) and 835  tokens (15\%) across 1,095,374 transactions (97\%).

\section{Ego Multi-Token Network}
We present the method for obtaining the ego multi-token network from the ego token transfer dataset. We also describe the process for extracting the network backbone, measuring users' and tokens' centralities, and balance scores.

\paragraph{\textbf{Network construction}}
The naive network representation of the dataset previously described is a \emph{multigraph}, such that its nodes correspond to externally owned accounts (EOAs) or contract accounts (CAs), and directed multi-edges between a pair of nodes represent the token transfer direction.
In this study, we consider a different representation based on the network containing multiple instances of EOAs and CAs for each traded token, called a \emph{multi-token network (MTN)}. Such a duplicated-node network representation is used in the context of the World Trade Network \cite{coquideWTN_2,ermannWTN}.
The MTN contains nodes representing pairs of user and token $(u,t)$, where $u\in [1,N_{u}]$ is the user index, and $t \in [1,N_{t}$] is the token index. Since we consider an ego network, users can be either an \emph{ego} or an \emph{alter}. In the MTN, an edge $i'\rightarrow i$, with $i' = (u',t)$ and $i = (u,t)$ two nodes of the network, represents the number of transactions in token $t$ sent by $u'$ and received by $u$. Possible edges are between egos or between alter and ego. Adjacency matrix elements are
\begin{equation}
    A_{ii'}=M^{t}_{u,u'}\delta_{tt'}
\end{equation}
with $M^{t}_{u, u'}$ the number of transactions in token $t$ sent by $u'$ and received by $u$. $\delta_{tt'}$ is the Kronecker delta function whose output is $1$ for $t=t'$ and $0$ otherwise. 
Due to the presence of different natures of tokens, edge weights are based on the number of transactions rather than the token's value.

\paragraph{\textbf{Backbone extraction}}
\label{sec:backbone}
Exploring MTN is not straightforward since its size is large and it contains disconnected islands. For this reason, we perform a backbone extraction method that permits us to maintain significant interactions. We use the multi-scale backbone extraction method from \cite{guell_bb}, which considers both the edge weights and directions in the measure of the edges' significance score.

Let $i=(u,t)$ and $i'=(u',t)$ be two nodes associated with a user $u$ and one of its economic partners $u'$ for token $t$. We denote by $k^{in}_{i}$ (resp. $k^{out}_{i}$) the number of distinct partners sending (receiving) token $t$ to (from) user $u$. If $u'$ is an outgoing neighbor of $u$ (i.e., a destination of transactions in token $t$ from $u$), the significance score associated to that link is

\begin{equation}
    S_{i'i} = A_{i'i}/\sum_{k}A_{ki}
\end{equation}
with $\sum_{k}A_{ki}$ the total number of transactions in token $t$ sent by $u$. 
Regarding an arbitrary significance threshold $\alpha \in [0,1]$, this link is kept if 
\begin{equation}
    (1-S_{i'i})^{k^{out}_{i}-1} < \alpha
\end{equation}
In the same way, the incoming link significance score is
\begin{equation}
    S_{ii'} = A_{ii'}/\sum_{k}A_{ik}
\end{equation}
and the condition to keep it in the backbone is 
\begin{equation}
    (1-S_{ii'})^{k^{in}_{i}-1} < \alpha
\end{equation}

The significance threshold value $\alpha = 0.001$ used in this study permits the contention of a backbone containing few nodes and edges with a higher density of connections.

\paragraph{\textbf{User and token centrality measures}}
The MTN representation permits centrality measures for both the users and the tokens. We present here the centrality metrics used and their economic interpretations.

\paragraph{PageRank and token accumulators.} The PageRank (PR) is an eigenvector-based centrality measure. The PR score associated with a node $i$ is the probability of finding a random walker at node $i$ after an infinite trip within the network. Since link direction in the MTN is associated with the token destination, PR probability associated with a doublet $(u, t)$, noted $p_{ut}$, is interpreted as the probability that token $t$ is accumulated by user $u$ within the network.

\paragraph{CheiRank and token spreader.} The CheiRank (CR) scores are the PageRank scores associated with the transpose graph (i.e., the graph with link directions reversed). Since the direction of links, CR score associated to the node $(u,t)$, noted $p^{*}_{ut}$, is interpreted as the probability token $t$ is spread from user $u$ within the network.

Tokens' and users' PR scores (resp. CR scores) are calculated as follows:
\begin{equation}
    p_{u}=\sum_{t}p_{ut},\qquad 
    p_{t}=\sum_{u}p_{ut}
    \label{eq:userprtokenpr}
\end{equation}
the score $p_{u}$ (resp. $p_{t}$) measures $u$ ($t$) accumulating capacity for all assets he is involved in (user involved in).


\paragraph{\textit{PageRank-CheiRank trade balance}}

We consider a measure based on PR and CR, allowing us to distinguish network nodes regarding their economic behavior. In \cite{coquideWTN_2}, the PageRank-CheiRank Trade Balance (PCTB) is proposed to measure countries' economic health. In the context of the MTN, we use the PCTB to measure both, the users' and tokens' economic behavior. We have
\begin{equation}
    B_{u} = \frac{p^{*}_{u} - p_{u}}{p^{*}_{u} + p_{u}},\qquad
    B_{t} = \frac{p^{*}_{t} - p_{t}}{p^{*}_{t} + p_{t}} 
\label{eq:userbaltokenbal}
\end{equation}
with $B_{u}$ and $B_{t}\in [-1,1]$, the PCTB score associated with the user and token, respectively. A positive (resp. negative) balance score is associated with an accumulating (spreading) behavior since the Pagerank contribution is higher (lower) than the CheiRank contribution.

We can measure the PCTB at the ($u$,$t$) scale with
\begin{equation}
    B_{ut}=\frac{p^{*}_{ut} - p_{ut}}{p^{*}_{u} + p_{u}}
    \label{eq:usertokenbal}
\end{equation}
The contribution of token $t$ to the total balance associated with user $u$.


\section{Results}
\subsection{Aggregated Ego Multi-Token Network}
\label{sec:staticnet}
The first analysis is based on the network representing the sum of token transfers for the whole period of the dataset from January 2018 to March 2023. We present structural metrics associated with the global network, the largest strongly connected component (SCC), and the backbone in \autoref{tab:stats}. The ego multi-token network (MTN) is sparse, the number of nodes and number of edges being of the same order of magnitude, and an edge density $d=1.8 \times 10^{-4}$.

\begin{table}
\caption{\label{tab:stats} Networks structural metrics: Number of nodes ($N$), users ($N_{u}$), tokens ($N_{t}$), links ($N_{l}$) and edge density ($d$).}
\centering
\renewcommand{\arraystretch}{1.5}
\begin{tabular}{|l|p{40pt}|p{40pt}|p{40pt}|p{40pt}|p{45pt}|}
\cline{2-6}
\multicolumn{1}{l|}{} &\cellcolor{gray!20} $N$ & \cellcolor{gray!20} $N_{u}$ &\cellcolor{gray!20}  $N_{t}$ &\cellcolor{gray!20}  $N_{l}$ &\cellcolor{gray!20}  $d$\\
\cline{1-6}
\cellcolor{black!20} \textbf{Global network}~~&\cellcolor{black!20} 83,570 &\cellcolor{black!20}  64,125 &\cellcolor{black!20}  603 &\cellcolor{black!20} 95,465 &\cellcolor{black!20} $ 1.8\times 10^{-4} $\\
\cline{1-6}
\cellcolor{gray!20} \textbf{Largest SCC}~~&\cellcolor{gray!20} 2.75 $(\%)$ &\cellcolor{gray!20} 3.6 $(\%)$ &\cellcolor{gray!20} 1 (ETH)&\cellcolor{gray!20} 5.44 $(\%)$ &\cellcolor{gray!20} $2\times10^{-2}$ \\
\cline{1-6}
\cellcolor{black!20} \textbf{Backbone}~~&\cellcolor{black!20} 1.85 $(\%)$ &\cellcolor{black!20} 0.9 $(\%)$ &\cellcolor{black!20} 21.4 $(\%)$ &\cellcolor{black!20} 1.5 $(\%)$ &\cellcolor{black!20} $1.2\times10^{-3}$\\
\cline{1-6}
\end{tabular}
\end{table}

\begin{figure}
    \centering
    \includegraphics[width=\linewidth]{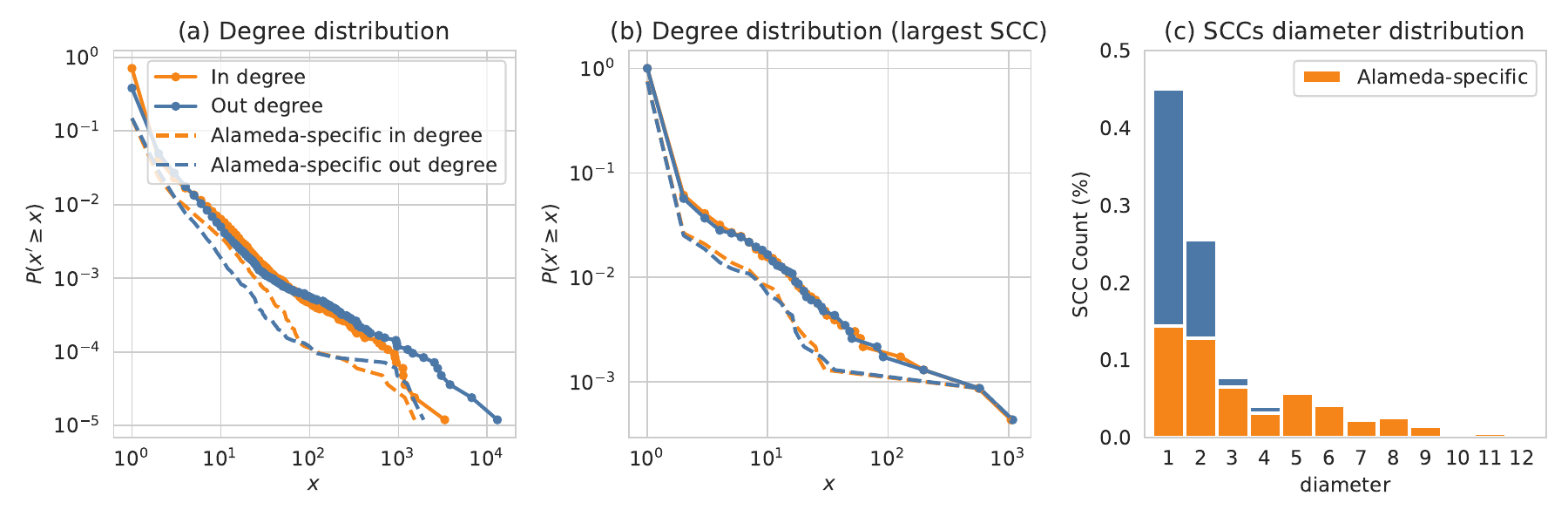}
    \caption{Node degree cumulative distribution regarding incoming (orange) and outgoing (blue) edges for (a) the global multi-token network, and (b) its largest SCC. (c) Diameter distribution for SCCs.}
    \label{fig:diststatic}
\end{figure}

The MTN shows a strong symmetry in incoming and outgoing degree distributions (see \autoref{fig:diststatic} (a)) in the case of all transactions and Alameda-specific ones.
This symmetry is also present in the largest SCC (see \autoref{fig:diststatic} (b)).
Due to the construction of the multi-token network, the existing SCC only contains users who share identical tokens. \autoref{fig:diststatic} (c) shows the distribution of the diameter of the 570 SCCs. $54\%$ of them contains Alameda accounts. Most of these SCCs have diameter $d\leq3$, indicating a large portion of dyadic interactions. These structures represent small groups of users sharing a small set of tokens. Alameda accounts are present in all SCCs with $d>5$. The largest SCC is related to ETH trades. It has a diameter $d=9$ and contains $\approx 3\%$ of all nodes ($\approx 4\%$ of all users), including Alameda and 11 other fund accounts.

\begin{figure}
    \centering
    \includegraphics[width=1\linewidth]{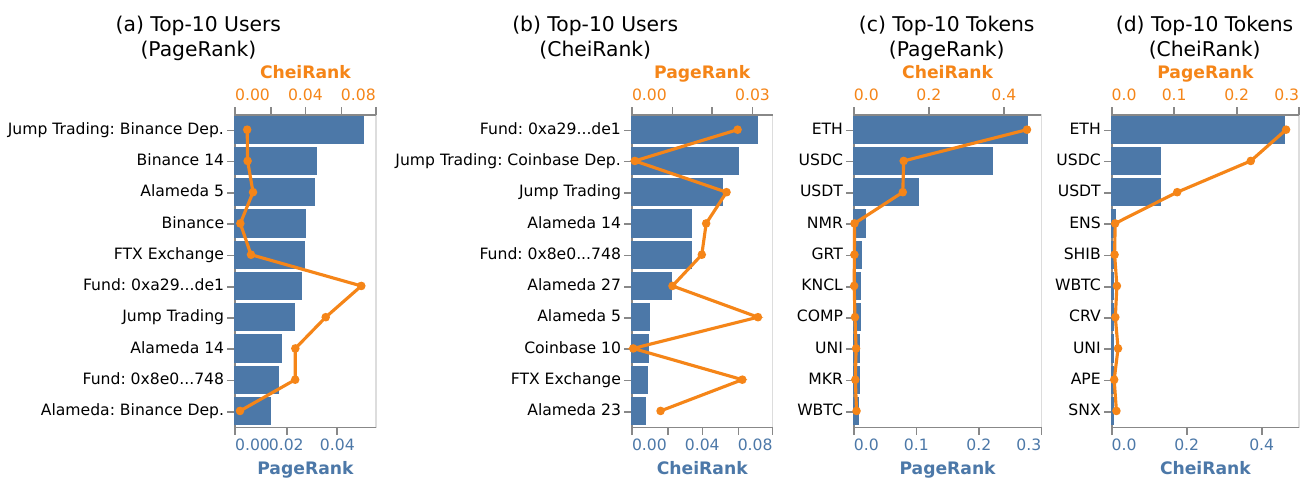}
    \caption{Top-10 users and tokens regarding PageRank and CheiRank scores}
    \label{fig:top_pagerank}
\end{figure}

We identify the top accounts and tokens and analyze their positions within the network. To do so, we list the top 10 accounts and tokens in terms of PageRank (PR) and CheiRank (CR) in \autoref{fig:top_pagerank}. Top users issued from PR and CR scores are mostly \emph{egos}, including \emph{Alameda and Jump Trading}. Among top \emph{alter} users, we find \emph{Binance and FTX}, centralized cryptocurrency exchange services. Scores associated with the top 10 users are spread relatively equally among them. There is a low similarity between the two tops based, as measured by a rank-biased overlap (RBO) score of $r=0.24$.
On the contrary, only a few tokens dominate the transfer network. Three of them, ETH (cryptocurrency), USDC, and USDT (stablecoins), concentrate most of the highest scores, while other tokens present a score $<1\%$. PR and CR top-10s of tokens show a stronger similarity ($r=0.67$).


\begin{figure}
    \vspace{-0.2cm}
    \centering
    \includegraphics[width=1\linewidth]{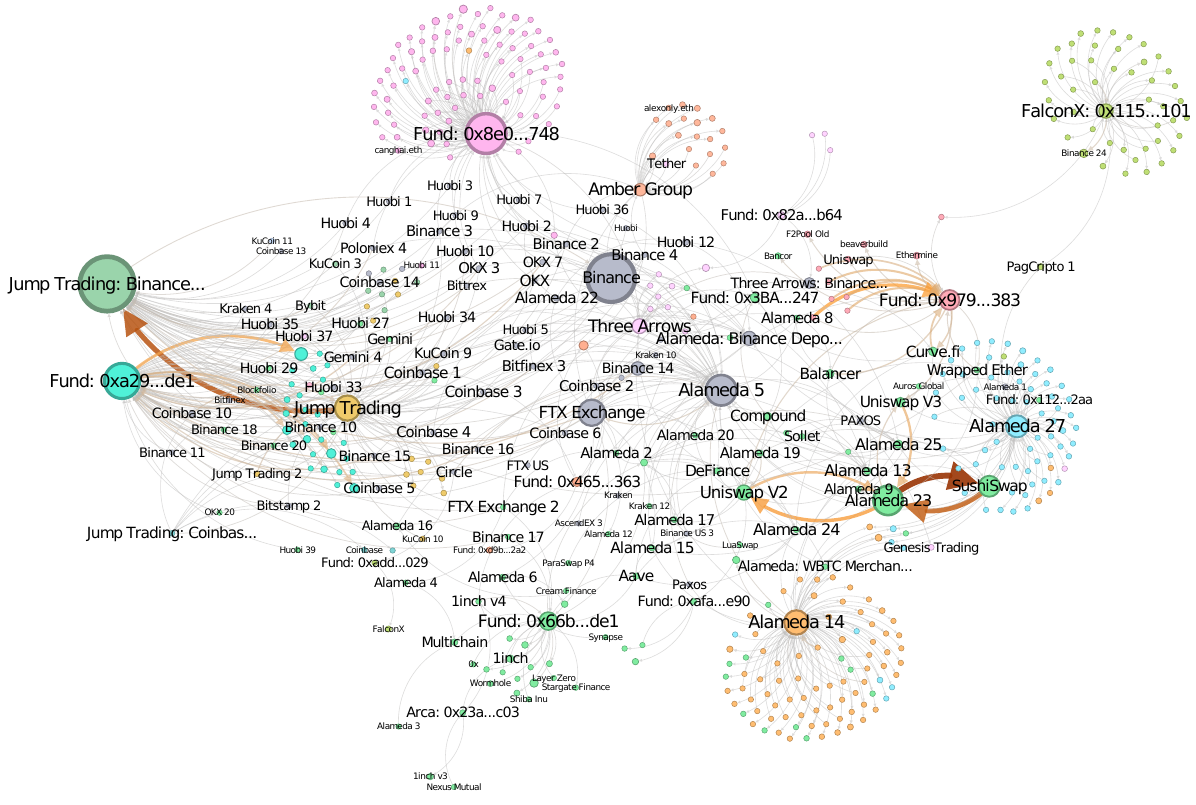}
    \caption{Visualization of the backbone network. Node color represents each cluster, while node size is proportional to the PageRank score.
    Link thickness corresponds to the number of transactions between nodes.}
    \label{fig:backbone}
    \vspace{-0.3cm}
\end{figure}

To analyze different groups of activities within the token transfer network, we applied the Louvain community detection algorithm, which grouped nodes into 16 clusters with a modularity score of 0.81. Given the network's sparsity, \autoref{fig:backbone} visualizes the backbone network, where the node colors represent the clusters.

The backbone network reveals a hub-and-spoke structure, with large fund accounts, notably \emph{Alameda and Jump Trading}, acting as central hubs. These accounts are highly connected to a diverse set of smaller accounts and contracts. The network is divided into distinct regions, with Alameda-related accounts dominating the right side and Jump Trading forming a separate region on the left.
This partitioning suggests distinct activity patterns and strategies between these entities, which we will explore further through specific cluster analysis.



\begin{itemize}
    \item \textbf{Cluster 5 \tikzcircle{90edff}{64a5b2}, 6 \tikzcircle{ffbd71}{b2844f} 7 \tikzcircle{ffb6ee}{b27fa6}, 14 \tikzcircle{ffb79b}{b2806c}, and
    15 \tikzcircle{bfdf75}{859c51}} are typical hub-and-spoke accounts. Funds such as \emph{Alameda 27, Alameda 14, Fund: 0x8e0...748, Amber Group, and FalconX: 0x115...101} play the role of local centers, interacting exclusively with customer accounts and forming star-like subgraphs in the backbone network.
    \item \textbf{Cluster 3 \tikzcircle{ffaab7}{b27680} and 16 \tikzcircle{80eca0}{59a570}} interact with a smaller group of nodes. They seem to be specialized in interactions with DeFi protocols.
    For instance, \emph{Alameda 23} (Cluster 16 \tikzcircle{80eca0}{59a570}) engages in a large number of token swaps with decentralized exchanges like \emph{Uniswap V2} and \emph{SushiSwap}. \emph{Fund: 0x979...383} (Cluster 3~\tikzcircle{ffaab7}{b27680}) receives tokens from numerous accounts, including other fund accounts like \emph{Alameda 8 and Three Arrows: Binance}, and conducts significant token exchanges like USDT and WETH with the \emph{Curve.fi} and \emph{Uniswap}.
    \item \textbf{Cluster 1 \tikzcircle{b7bbcb}{80828e}, 8 \tikzcircle{4ff1d8}{37a897} and 13 \tikzcircle{efcb6b}{a78e4a}} seem to include fund and exchange. The main fund accounts in these clusters, \emph{Alameda 5, Fund: 0xa29...de1 and Jump Trading}, respectively, are connected to off-chain exchange companies such as \emph{Binance, Coinbase, and Huobi}.
    Notably, \emph{Jump Trading} sent tokens to \emph{Jump Trading: Binance Deposit} (Cluster 2 \tikzcircle{9ad4ab}{6b9477}) while \emph{Alameda 5} (Cluster 1 \tikzcircle{b7bbcb}{80828e}) transferred USDC to the \emph{FTX Exchange}.

\end{itemize}

\subsection{Snapshot Network Analysis}
\label{sec:snapshot}

\begin{figure}
    \centering
    \includegraphics[width=1\linewidth]{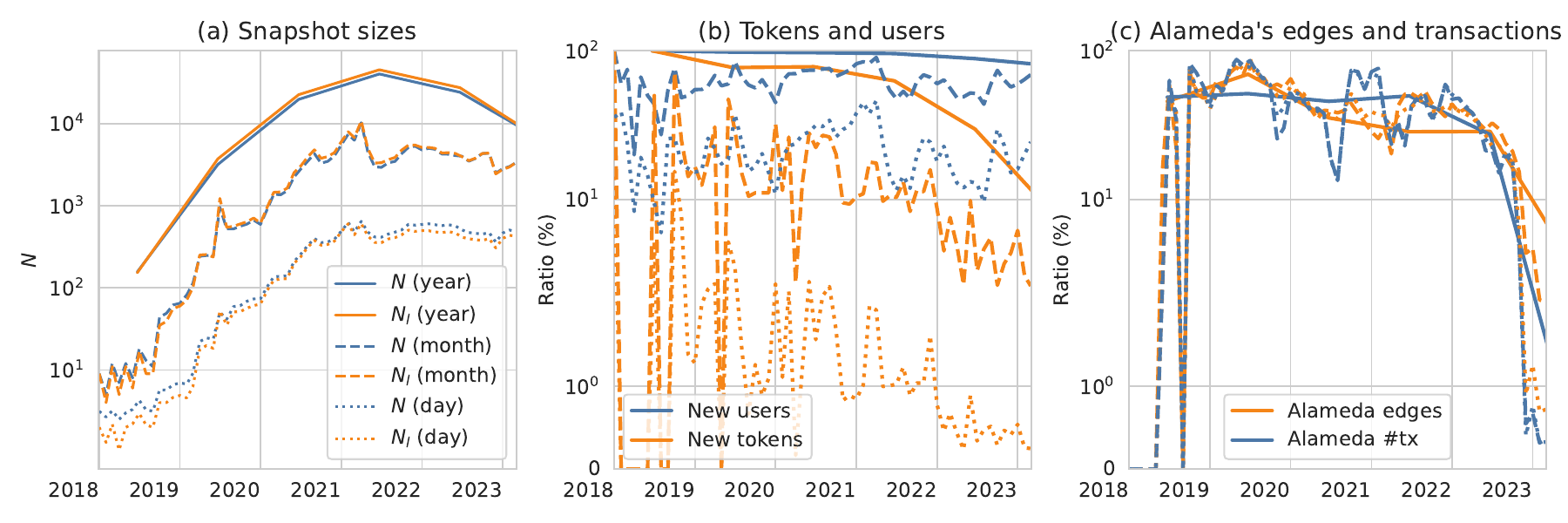}
    \caption{Evolution of snapshot sizes and compositions: (a) number of nodes (blue) and edges (orange), (b) percentage of new users (blue) and new tokens (orange), and (c) number of edges (orange) and transactions (blue) Alameda is involved. Yearly (solid lines), monthly (dashed lines) and daily (dotted lines).}
    \label{fig:snap1}
\end{figure}

We analyze the time evolution of the ego multi-token network (MTN). The evolution of the number of nodes and edges at year, month, and day resolution is presented in \autoref{fig:snap1} (a). Note that certain days have no transactions; therefore, daily results are averaged to produce a monthly value, thus reducing significant fluctuations. Similar to the aggregated MTN in \autoref{sec:staticnet}, snapshots show a ratio $N/N_{l}$ close to $1$. Every time-resolution shows rapid growth in both nodes and edges. \autoref{fig:snap1} (b) shows the time evolution of the ratio of new incoming users and new exchanged tokens in MTN snapshots. It appears that network change in size is primarily due to new incomers, while the ratio of new exchanged tokens decreases over time.
\autoref{fig:snap1} (c) presents the temporal evolution of the number of Alameda's outgoing and incoming edges and the associated number of transactions. Its entrance into the MTN was observed around June 2018 (low values). While the evolution of the number of edges associated with Alameda remains stable from 2018 to mid-2022, the evolution of the number of transactions shows strong variations. Two local minima were observed before 2021, and a large peak was observed during the first half of 2021. During Alameda's bankruptcy in November 2022, both the transitions and edges abruptly decreased.

\begin{figure}
    \centering
    \includegraphics[width=1\linewidth]{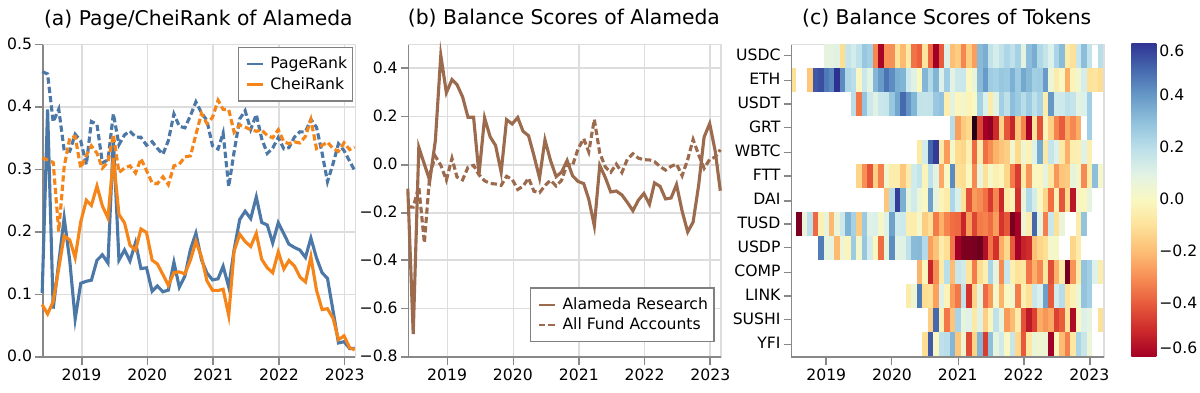}
    \caption{Temporal evolution of Alameda Research's centralities of (a) PageRank and CheiRank scores, (b) PageRank-CheiRank Trade Balance scores for Alameda Research compared to all fund accounts, and (c) balance scores for the top-10 tokens for Alameda Research. Snapshots with month resolution are considered.}
    \label{fig:balance}
\end{figure}

We focus on Alameda Research accounts to analyze their monthly temporal evolution and activities, particularly leading up to key events such as Alameda's eventual collapse in November 2022.
\autoref{fig:balance} (a--b) presents the evolution of PageRank (PR) and CheiRank (CR) scores as well as the PageRank-CheiRank Trade Balance scores (PCTB) for the 27 accounts owned by Alameda and compares these scores with ones associated with other fund accounts. While PR and CR scores associated with other fund accounts are stable for the entire period, Alameda's scores present moments of burst around July 2019, October 2020, and September 2021 (see \autoref{fig:balance} (a)). Additionally, a strong decrease in scores is observed at the end, starting from November 2022 and lasting for the rest of the period covered by the dataset. The time evolution of the PCTB scores is presented in the \autoref{fig:balance} (b). While the trading behavior of other fund accounts is balanced (PCTB score around 0) and stable with time, Alameda's PCTB score is globally decreasing with time. Alameda Research passes from a token spreader behavior ($t<$ November 2020) to a token accumulator behavior until October 2022. Around Alameda Research's bankruptcy (November 2022), we observe a succession of high token spreader and token accumulator behavior.

\autoref{fig:balance} (c) displays the PCTB scores over time for the tokens in the top 10 for PageRank and CheiRank of the Alameda accounts.
These scores permit us to infer token-specific Alameda's trading behavior. We observe that behaviors are token-dependent. Regarding ETH (cryptocurrency), Alameda presents a token spreader behavior for almost all periods, except in 2019, and March 2023, whose behavior is more related to accumulation. Regarding USDC and USDT (stablecoins), Alameda plays a different role in the network. In the case of USDC, there is a transition from a spreading phase (2020-2021) to an accumulating phase (2021-2022), whereas its behavior regarding USDT is more stable and related to a spreader. GRT, USDP, TUSD, and DAI are the tokens Alameda accumulates the most (highly negative PCTB values).
Interestingly, our analysis shows that Alameda's behavior regarding FTX's native token, named FTT, is accumulating, such as what CoinDesk reported online\footnote{\url{https://www.coindesk.com/business/2022/11/02/divisions-in-sam-bankman-frieds-crypto-empire-blur-on-his-trading-titan-alamedas-balance-sheet/}}.

\section{Conclusion and Future Work}
\label{sec:conclusion}
Our analysis of Alameda Research's ego token transfer network and related accounts shed light on the complexity of token transactions within the DeFi ecosystem, leading to insights into transaction dynamics and token flows. We show that the aggregated multi-token transfer network exhibits significant structural properties, such as solid symmetry of in- and out-degree distributions, the presence of dominant small and tightly connected groups of users, and star-like structures centered on major players like Alameda and Jump Trading. The use of PageRank and CheiRank centralities permits the identification of top accounts and tokens. The analysis shows that few dominant tokens concentrate on DeFi activities. By considering a temporal graph approach, we demonstrate Alameda's fluctuating network activity, with notable shifts in token distribution and accumulation leading up to its bankruptcy.

Further work based on a temporal community detection approach and anomaly detection will provide deeper insights into important transactions that have led Alameda to bankruptcy. We will also consider stream graph representations, which give fine-grained descriptions of the transaction dynamics.

%
%
\bibliographystyle{spmpsci} 
\bibliography{refs} 
\end{document}